\documentclass[12pt,preprint]{aastex}
\usepackage{natbib}
\shorttitle{Interferometric mapping of magnetic fields}
\shortauthors{P. C. Cortes, R. M Crutcher, B. C. Matthews}

\begin{document}

\title{Interferometric Mapping of Magnetic Fields: NGC2071IR }
\author{Paulo C. Cortes}
\affil{Departamento de Astronom\'ia, Universidad de Chile, Casilla 36-D Santiago, Chile }
\author{Richard M. Crutcher}
\affil{Astronomy Department, University of Illinois at
    Urbana-Champaign, IL 61801, U.S.}
\author{Brenda C. Matthews}
\affil{Herzberg Institute of Astrophysics, National Research Council of Canada, Victoria, BC, 
V9E 2E7, Canada}

\begin{abstract}
We present polarization maps of NGC2071IR 
from thermal dust emission
at 1.3 mm  and from CO J=$2 \rightarrow 1$ line emission.
The observations were obtained using the Berkeley-Illinois-Maryland
Association array in the period 2002-2004.
We detected dust and line polarized emission from NGC2071IR that
we used to constrain the morphology of the magnetic field.
From  CO J=$2 \rightarrow 1$  polarized  emission we found evidence
for a magnetic field in the powerful bipolar outflow present
in this region. We calculated a visual extinction 
$A_{\rm{v}} \approx 26$
mag from our dust observations. This result, when compared with
early single dish work, seems to show that 
dust grains emit polarized radiation efficiently at higher densities than
previously thought. Mechanical alignment by the outflow is proposed
to explain the polarization pattern observed in NGC2071IR, which
is consistent with the observed flattening in this source.
\end{abstract}


\section{Introduction}

Without a doubt, the star formation process is one of most complicated
astrophysical phenomena yet to be explained. The diversity of dynamically
significant parameters makes it, still, an unsolved problem.
One of the common features observed in star forming regions
are molecular outflows. These energetic gas flows can reach supersonic
motions at velocities up to $\sim 50$ km s$^{-1}$ and shock
front temperatures up to $\sim 2000$ K. 
Outflows are not well understood, despite considerable advances
toward the understanding of the flow morphology, shock propagation,
and wind chemistry; still the basic driving mechanisms for these flows
are not well understood. Current models
(see \citet{Bachiller1996} for a review) predict the eruption 
of jets from the poles of protostars at certain stages of star
formation. These models require alignment of the gas with a magnetic
field which collimates the gas forming the jet. Therefore, observing magnetic
fields in star forming regions is critical to understand the
complexity of molecular outflows.

Magnetic field observations are divided into measurements of the Zeeman
effect (in order to obtain the magnetic field strength in the line
of sight), and linear polarization observations of dust and spectral-line emission.
Polarization of dust emission is believed to be perpendicular to the magnetic field
under most conditions \citep{Lazarian2003}; hence, polarization of dust emission
has been used as a major probe for the magnetic field geometry.
In order to efficiently map the polarization of dust emission and infer
information about the magnetic field morphology, high resolution
observations are required. The BIMA millimeter interferometer
has been used previously to obtain high-resolution polarization maps
in several star forming cores \citep{Rao1998,Girart1999b,Lai2001,
Lai2002,Lai2003}.
Spectral line linear polarization has been suggested to arise
from molecular clouds under
anisotropy conditions \citep{Goldreich1981}. The prediction suggests
that a few percent of
linearly polarized radiation should be detected from molecular clouds and
circumstellar envelopes in the presence of a magnetic field. It is also
predicted that the molecular line polarization will be either parallel
or perpendicular to the magnetic field, depending on the angles between
the line of sight, the magnetic field, and the anisotropic excitation
direction \citep{Goldreich1982}. This
process is known as the Goldreich - Kylafis effect.

One of the most powerful outflows known to date is observed in NGC2071IR
 \citep{Bally1982}. This outflow has been extensively studied (see 
$\S 2$), however, its origin is still not clear. The outflow
orientation has been measured by molecular line observations of HCO$^{+}$
\citep{Girart1999b} and CO $J=2 \rightarrow 1$ \citep{Moriarty-Schieven1989},
to be $\sim 40^{\circ}$ to $50^{\circ}$.
The orientation of the outflow remarkably agrees with polarization observations
of dust emission by \citet{Matthews2002}. 
If a magnetic field is responsible for the collimation of outflows,
we should be able to trace the field orientation in the outflow using
the techniques previously described.

In order to study the magnetic field 
we mapped NGC2071IR with the BIMA array.
We measured continuum polarization at 1.3 mm and
CO $J=2\rightarrow 1$ line polarization obtaining high resolution
interferometric maps for both measurements.

The remainder of this paper is divided in five major sections. Section 2 reviews
information about the source, section 3 describes the observation
procedure.
Section 4 presents the results, section 5 gives the discussion,
and section 6 the conclusions and summary.

\section{Source Description}

\begin{figure}
\label{1}
\figurenum{1}
\includegraphics[scale=0.7]{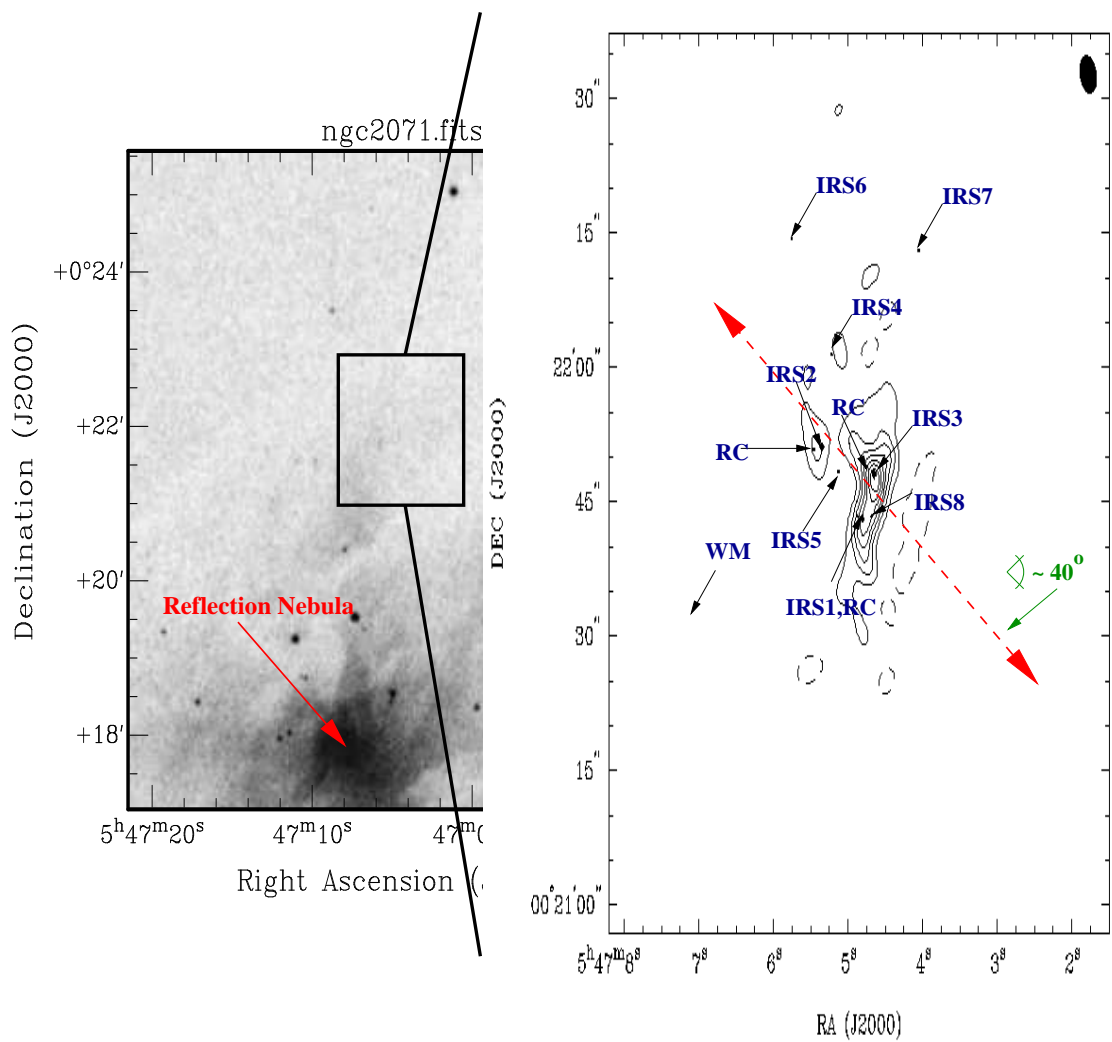}
\caption{The Figure shows a schematic map of the NGC2071IR region. The
contours are a qualitative representation of the main core mapped. The
IRS sources are shown according to the numbering given by \citet{Walther1993}.
Some of the IRS sources are out of the main area mapped by Figure \ref{NGC2071con}.
The RC sources are the radio continuum and possible background sources
reported by \citet{Snell1986}. The WM source represent the H$_{2}$O maser
reported by \citet{Schwartz1975}. The NGC2071 nebula is also shown.
The bipolar outflow is shown by the
arrow with an overall P.A. of 40$^{\circ}$.}
\label{RPpic}
\end{figure}

NGC2071 is a well-known optical reflection nebula located in the dark cloud
Lynds 1630, which is part of the Orion complex. NGC2071IR, located
4$^{\prime}$ north from the reflection nebula, is a well
studied region with  a distance of 390 pc
\citep{Anthony-Twarog1982}
and a size $\sim$ 30$^{\prime \prime}$. 
Figure \ref{RPpic} shows a qualitative picture
of the NGC2071IR region where all the main sources are identified. 
The NGC2071IR region is shown as a zoom in from a bigger region
which includes the NGC2071 optical reflection nebula.
The NGC2071IR region
has been resolved into eight distinct near-infrared sources
\citep{Walther1993} having a total luminosity of 520 $L_{\sun}$
\citep{Butner1990}. Two of them
(IRS 1 and IRS 3) are associated with 
5 GHz radio continuum emission from several point sources
\citep{Snell1986}; several of these radio continuum sources
 may be extragalactic background objects.
IRS 1 dominates
the emission at near-infrared wavelengths, while IRS 3 dominates at longer wavelengths \citep{Snell1986}.
H$_{2}$O maser emission has been detected toward NGC2071IR
\citep{Schwartz1975,Pankonin1977,Campbell1978}, particularly at IRS 1 and
IRS 2, which seems to indicate the presence of substantial columns of
dust-laden, warm (300-1000 K), dense gas ($10^{8}-10^{10}$ cm$^{-3}$).
A 1720 MHz OH maser,  located between IRS 1 and IRS 2,
coincides with an H$_{2}$O maser. 

A powerful molecular outflow
was found in NGC2071IR \citep{Bally1982}; the outflow has been
mapped in CO  \citep{Snell1984,Moriarty-Schieven1989}, in CS \citep{Zhou1991},
and  SO, SiO, and HCO$^{+}$ \citep{Chernin1992,Chernin1993,Garay2000,Girart1999a}. The origin of this outflow is attributed to IRS 1; the best evidence for this comes from H$_{2}$ observations of \citet{Aspin1992}. We show the outflow direction in Figure \ref{RPpic}.
However, observations
from \citet{Garden1990} show elongated molecular emission associated
with IRS 3, which is only $\sim 10^{\prime\prime}$ from IRS 1. In addition, high resolution images of radio continuum
emission show elongated emission which is coincident with both infrared sources
\citep{Torreles1998b,Smith1994,Snell1986}.
Polarimetry observations have also been made of NGC2071IR. \citet{Walther1993} made high
resolution $K$ band imaging polarimetry over the whole cluster. IRS 1 and
IRS 3 are highly polarized ($> 30 \%$ for IRS1 and $\sim 20\%$ for IRS 3),
while IRS 2, IRS 4 and IRS 6 show low polarization.
\citet{Matthews2002} measured continuum linear polarization toward NGC2071IR
at 850 $\mu$m using the James Clerk Maxwell Telescope (JCMT).
They found a polarization pattern which is ordered and
qualitatively similar to other star forming regions, such as OMC-1
\citep{Schleuning1998}. However, they interpreted their polarization
pattern differently from \citet{Schleuning1998}. The \citet{Matthews2002}
polarization maps do not show the pinch seen in OMC-1; they suggested that
the magnetic field morphology is
inconsistent with a  dynamically significant magnetic field threading
the NGC2071IR core with a
hourglass shape. They also concluded that
the emission is dominated by dust, discarding contamination
from the CO $J=3\rightarrow2$ line from the powerful bipolar outflow that
originates out of the core. Applying the Chandrasekhar-Fermi method
 \citep{Chandrasekhar1953}, they estimated a magnetic field strength of 56 $\mu$G.

\section{Observation Procedure}

We observed NGC2071IR between October 2002 and May 2004, mapping the continuum
emission at 1.3 mm and the CO $J=2 \rightarrow 1$ molecular line (at 230 GHz).
Four tracks were obtained with the BIMA array in C configuration.
We set  the digital correlator in mode 8
to observe both the continuum and the  CO $J=2 \rightarrow 1$ line
simultaneously. The 750 MHz lower side band was combined
with  700 MHz from the upper side band to
map the continuum emission, leaving a
50 MHz window for the CO line observation (at a resolution of 1.02 km s$^{-1}$).
Each BIMA telescope has a single receiver, and thus the two polarizations
were observed sequentially.  A quarter wave plate to select either right (R)
or left (L) circular polarization was alternately switched into the signal path
ahead of the receiver.  Switching between polarizations was sufficiently
rapid (every 11.5 seconds) to give essentially identical uv-coverage.
Cross-correlating the R and L circularly polarized signals
from the sky gave RR, LL, LR, and RL for each interferometer baseline,
from which maps in the four  Stokes parameters were produced.
The source 0530+135 was used as phase calibrator for NGC2071IR.
The instrumental polarization was calibrated by observing
3C279,
and the ``leakages" solutions were calculated from this observation.
We used the same calibration procedure
described by \citet{Lai2001}.

The Stokes images I, U, Q and V were obtained by Fourier transforming
the visibility data using natural weighting.
The MIRIAD \citep{MIRIAD1995} package was used for data reduction.
Both sources are close to the equator; therefore, we expect strong
sidelobes in the beam pattern. We followed \citet{Chernin1995}, who
observed NGC2071IR with the BIMA array, 
and imaged only out to 20$^{\prime \prime}$
radius, due to the strong sidelobes.

\section{Observational Results}

\subsection{1.3 mm Continuum}

Figure~\ref{NGC2071con} shows our 1.3 mm continuum intensity and polarization 
maps; the  synthesized
 beam size is $4.4^{\prime \prime} \times 3.2^{\prime \prime}$ with a 
beam P.A. of 20.9$^{\circ}$. 
The contours represent
Stokes I with a peak emission of 0.3 Jy beam$^{-1}$. The total flux integrated
over a $20^{\prime \prime}\times 30^{\prime \prime}$ box is $S_{\nu,int}=1.87$ Jy.
This box defines our core in this region.
The gray scale in Figure~\ref{NGC2071con} corresponds to
polarized flux ($\sqrt{Q^{2} + U^{2}}$) which by definition has the
same units as the Stokes I emission, in this case the units are
Jy beam$^{-1}$. The line segments show the P.A.
and fractional polarization (which corresponds to the length
of the line segments). The polarization results shown are 
$3\sigma$ or better, which means that all results below the $3\sigma$
threshold are cutoff from the map\footnote{All our polarization results
at the center of our core were in the $2\sigma$ level. Therefore, are not
shown}.

The Stokes I map resolves
two components that are associated with some of the infrared sources reported
by \citet{Walther1993}.
The first clump is located in the central part of the map
corresponding to  an elongated and flattened structure from north to south
covering $\sim 30^{\prime \prime}$ in declination; this structure  
breaks up into smaller clumps with higher angular resolution.
This clump seems to be  associated with IRS 1, IRS 3, and IRS 8.
The second clump, associated with the IRS 2 source, is located to the east
of the first clump and is less significant.
The rest of the infrared sources do not show significant continuum emission
at 1.3 mm. \cite{Scoville1986}  observed the same region with the OVRO array
in 2.7 mm continuum obtaining
a peak flux of 0.17 Jy. Their map shows a core of circular morphology
which does  not show the same elongated morphology that our observations do
(see Figure \ref{NGC2071con}),
but this may be explained in terms of their lower resolution
($7^{\prime \prime}$) and sensitivity.

Most of the polarized emission is associated with
the central and strongest clump in three distinctive regions.
The northern region shows a
mean P.A. of $63.4^{\circ} \pm 7.8^{\circ}$, which
is polarization orthogonal to the major axis of the core.
The southern region 
shows a mean P.A. of $21.7^{\circ} \pm 7.2^{\circ}$, 
while the central part shows a
mean P.A. of $-15.6^{\circ} \pm 9.0^{\circ}$. Clearly, the direction
of the polarization changes significantly over the elongated clump. 
The continuum peak presents
no polarization at all (even at the $2\sigma $ level). This strong
depolarization could be produced by the dust in the south 
producing polarization along the major axis and the dust in the north 
producing polarization along the minor axis of the core, which contributes
roughly equally near
the continuum peak position, resulting in essentially no net polarized flux.
Table 1 shows P.A. and fractional polarization for NGC2071IR.

\subsection{CO $J=2 \rightarrow 1$}

\subsubsection{Description}

\begin{figure}
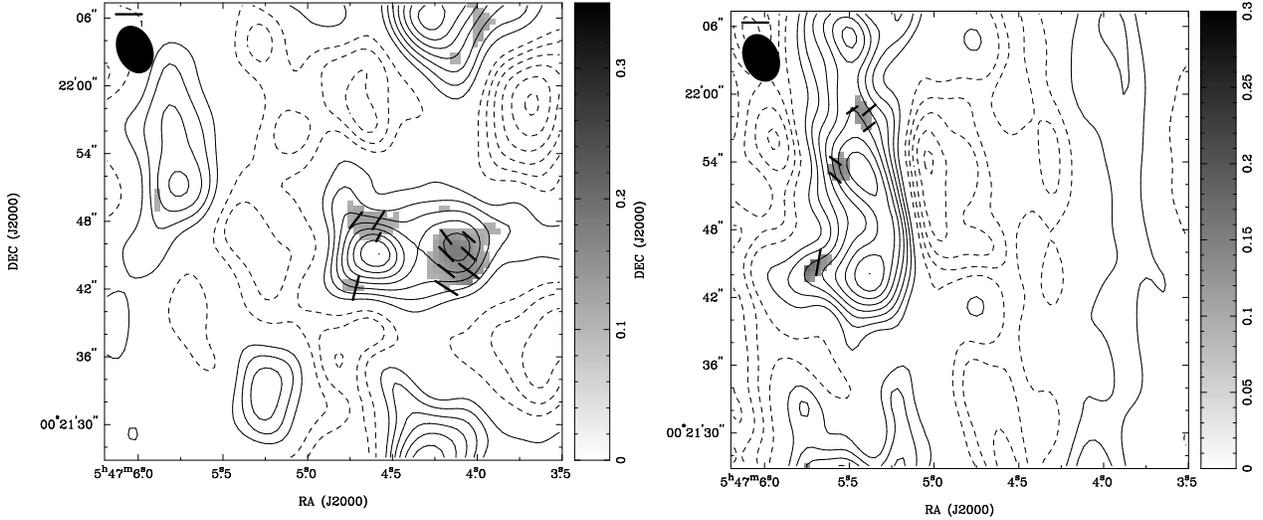

\label{3}
\figurenum{3}
\includegraphics[angle=-90,scale=0.35]{f3a.ps}
\includegraphics[angle=-90,scale=0.35]{f3b.ps}
\caption{The two maps represent
velocity averaged CO J=$2 \rightarrow 1$ polarized emission from NGC2071IR.
The panel shows the red-shifted emission (left panel) and blue-shifted
emission (right panel). The red-shifted emission was averaged over
v$_{lsr}=10.0$ to 23.0 km s$^{-1}$ in velocity.
The contour map represents Stokes I emission at levels:
-3.0, -2.4, -1.8, -1.0, -0.6, 0.6, 1.0, 1.8, 2.4, 3.0, 3.6,
and 4.2 Jy beam$^{-1}$ with a maximum of 4.8 Jy beam$^{-1}$
and noise level $\sigma=0.28$ Jy beam$^{-1}$.
The blue-shifted emission (right panel) was averaged over
v$_{lsr}=-7.8$ to -3.7 km s$^{-1}$ in velocity.
The contour map represents Stokes I emission at levels:
-4.1, -3.3, -2.4, -1.6, -0.8, 0.8, 1.6, 2.4, 3.3, 4.1, 4.9,
 and 5.7 Jy beam$^{-1}$ with a maximum of 6.5 Jy beam$^{-1}$ and
a noise level of $\sigma=$ 0.3 Jy beam$^{-1}$.
The gray scale represents $3\sigma$ polarized emission, which
is also in Jy beam$^{-1}$.
The length of the line segments are a representation of the fractional
polarization with scale of 0.086 show by the bar at the upper left corner.
The beam size is $4.4^{\prime \prime} \times 3.2^{\prime \prime}$ in
both maps.
}

\label{NGC2071co}
\end{figure}

Figure~\ref{NGC2071co} shows two panels with CO $J=2 \rightarrow 1$ maps
averaged in velocity. We used the velocity intervals: v$_{lsr}=-7.8$ to
-3.7 km s$^{-1}$ and v$_{lsr}=+10.0$ to  +23.0 km s$^{-1}$. Over 
v$_{lsr}$=-3.7 to +10.0 km s$^{-1}$ the polarized line flux is extremely weak.
These spectral-line maps are heavily affected by missing zero and
short-spacing visibility data. Nevertheless, they give at least
qualitative information. The NGC2071IR
molecular outflow can be seen in the blue-shifted and red-shifted maps;
the emission in these maps is consistent with previous observations
\citep{Moriarty-Schieven1989,Scoville1986}. The higher resolution in our maps allows
us to see more detailed structure in the outflow, particularly the
elongated structure in the blue-shifted lobe which is not resolved by
\citet{Moriarty-Schieven1989}. Also, the detection of polarized line
emission provide direct evidence for the presence of a magnetic field in the outflow.

\subsubsection{Polarized emission from the red-lobe}

The red-shifted lobe shows predominantly
 one region
of polarized emission, located between $\alpha=5^{h}47^{m}4.8^{s}$ to
$5^{h}47^{m}3.8^{s}$, and
$\delta=00^{\circ}21^{\prime}40^{\prime \prime}$ to
$00^{\circ}21^{\prime}52^{\prime \prime}$. The polarized
emission shows two distinctive orientations. The eastern-most region
has a mean P.A.  of  -31.5$^{\circ} \pm 8.7^{\circ}$
and a mean fractional polarization of 0.04$ \pm 0.01$;
the western region has a mean P.A. of 48.5$^{\circ} \pm 7.9^{\circ}$
and a mean fractional polarization of  0.05$ \pm 0.01$. This is a difference
in P.A. of $\Delta \phi \approx 80^{\circ}$ and is consistent with orthogonal
polarizations.

\subsubsection{Polarized emission from the blue-lobe}

The blue-shifted emission has three distinct regions of polarized emission,
which are located in the north-south direction at the same right ascension  ($\alpha \approx 5^{h}47^{m}5.5^{s}$).
The first region is centered at $\delta=00^{\circ}21^{\prime}58^{\prime \prime}$
 with mean P.A. of -50.2$^{\circ} \pm 9.2^{\circ}$ and mean fractional
polarization $0.04 \pm 0.01$. The second region is centered at
$\delta=00^{\circ}21^{\prime}53^{\prime \prime}$ with a mean
P.A. of $53.6^{\circ} \pm 8.9^{\circ}$ and mean fractional polarization
of $0.03 \pm 0.01$. The third region is centered at
$\delta=00^{\circ}21^{\prime}45^{\prime \prime}$ with a mean P.A. of
$9.8^{\circ} \pm 8.1^{\circ}$ and fractional polarization of
$0.07 \pm 0.01$. The first two regions have
a difference in P.A. of $\Delta \phi \sim 100^{\circ}$,
again consistent with orthogonal polarization.

\subsubsection{Polarized emission comparison}

The P.A. for the polarized emission in both lobes have similar
values. These values are orthogonal to each other and in agreement
with the prediction. However, this creates an ambiguity in the 
interpretation of the polarization. 
The outflow in NGC2071IR  has P.A. $\sim$ 40$^{\circ}$ to 50$^{\circ}$
\citep{Moriarty-Schieven1989,Girart1999a}, which is also schematically shown
by \citet{Matthews2002} in their maps. Our CO polarized emission presents P.A.
which are either parallel or perpendicular to the outflow direction
in the plane of the sky. From this result, the
projection of the magnetic field in the plane of the sky could be either
parallel or perpendicular to the outflow direction. Current models for
outflow suggest collimation by magnetic fields. Therefore, we believe that
the most plausible interpretation is a magnetic field, which is along 
the outflow.

\section{Discussion}

We centered
our observation at the same coordinates used by \citet{Matthews2002},
who observed at 850 $\mu$m in the continuum using the JCMT.
Their polarization map shows a uniform pattern over the core which
seems to be aligned with the outflow direction (as indicated in their maps).
Our continuum polarization results seem to agree reasonably well with
the single dish data \citep{Matthews2002}, except for our central region, 
where our results appear to be orthogonal to theirs. 
The strength of the polarized emission
is also  in agreement; their mean percentage polarization at
the core is about 5\% while our mean percentage polarization is about
6\% (see Table \ref{Table1.3}).

\begin{figure}
\label{2}
\figurenum{2}
\includegraphics[angle=-90,scale=0.7]{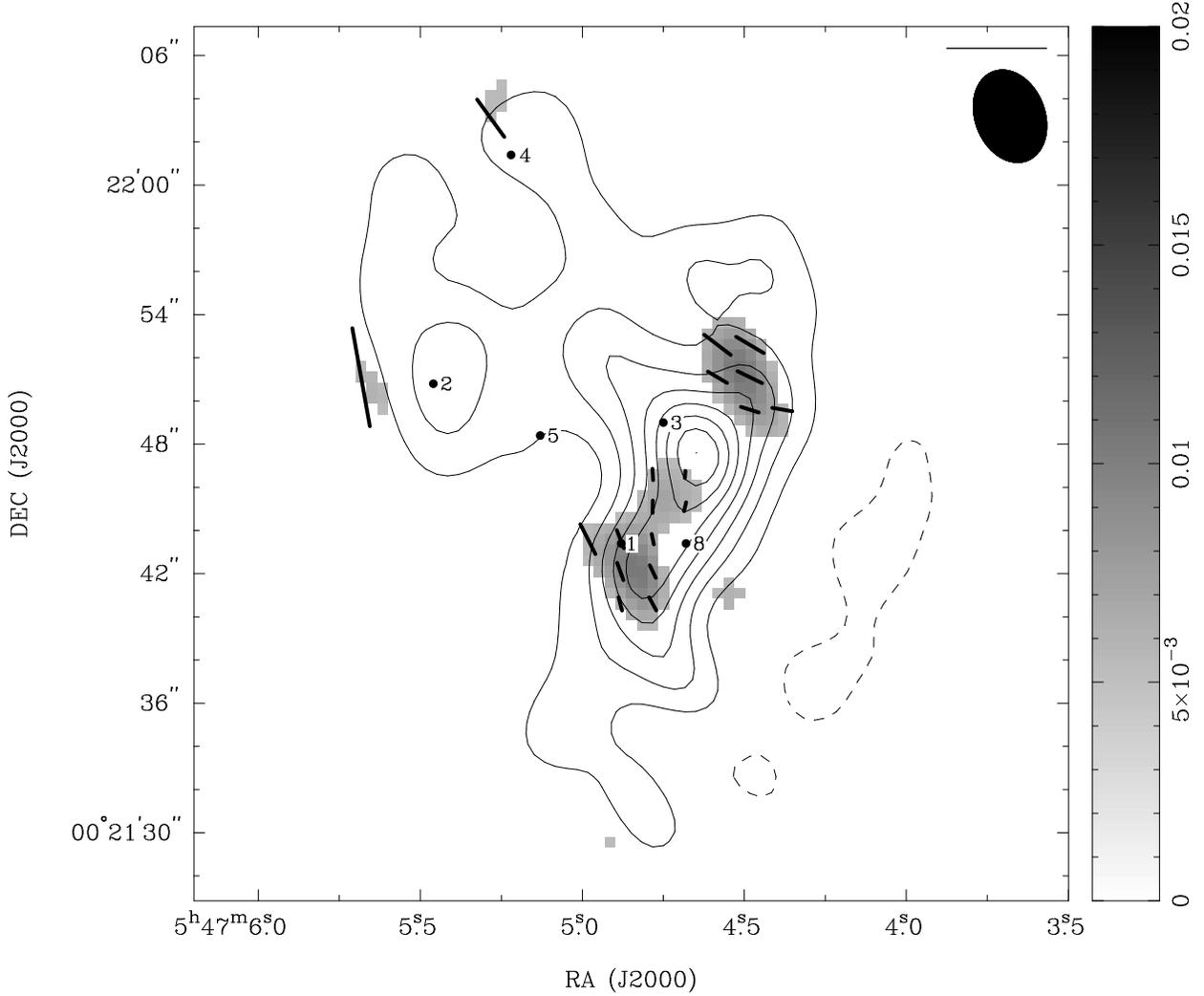}
\caption{Polarization map of NGC2071 at 1.3 mm. The contours represent
the Stokes I emission at  -0.03, 0.03, 0.07, 0.1, 0.14, 0.17, 0.21,
and 0.24 Jy beam$^{-1}$. The pixel gray scale shows $3\sigma$ polarized
intensity ($\sqrt{Q^{2} + U^{2}}$) for
the dust continuum emission also in Jy beam$^{-1}$. The line segments
are the polarization map, the length represents fractional polarization. 
The $\bullet$ symbols represent the
IRS sources. The rms in the Stokes I map is $\sigma=0.006$ Jy beam$^{-1}$.
The scale is set by 0.3 in fractional polarization shown by the
bar at the upper right corner of the map.
}
\label{NGC2071con}
\end{figure}

\begin{figure}
\label{4}
\figurenum{4}
\includegraphics[angle=-90,scale=0.7]{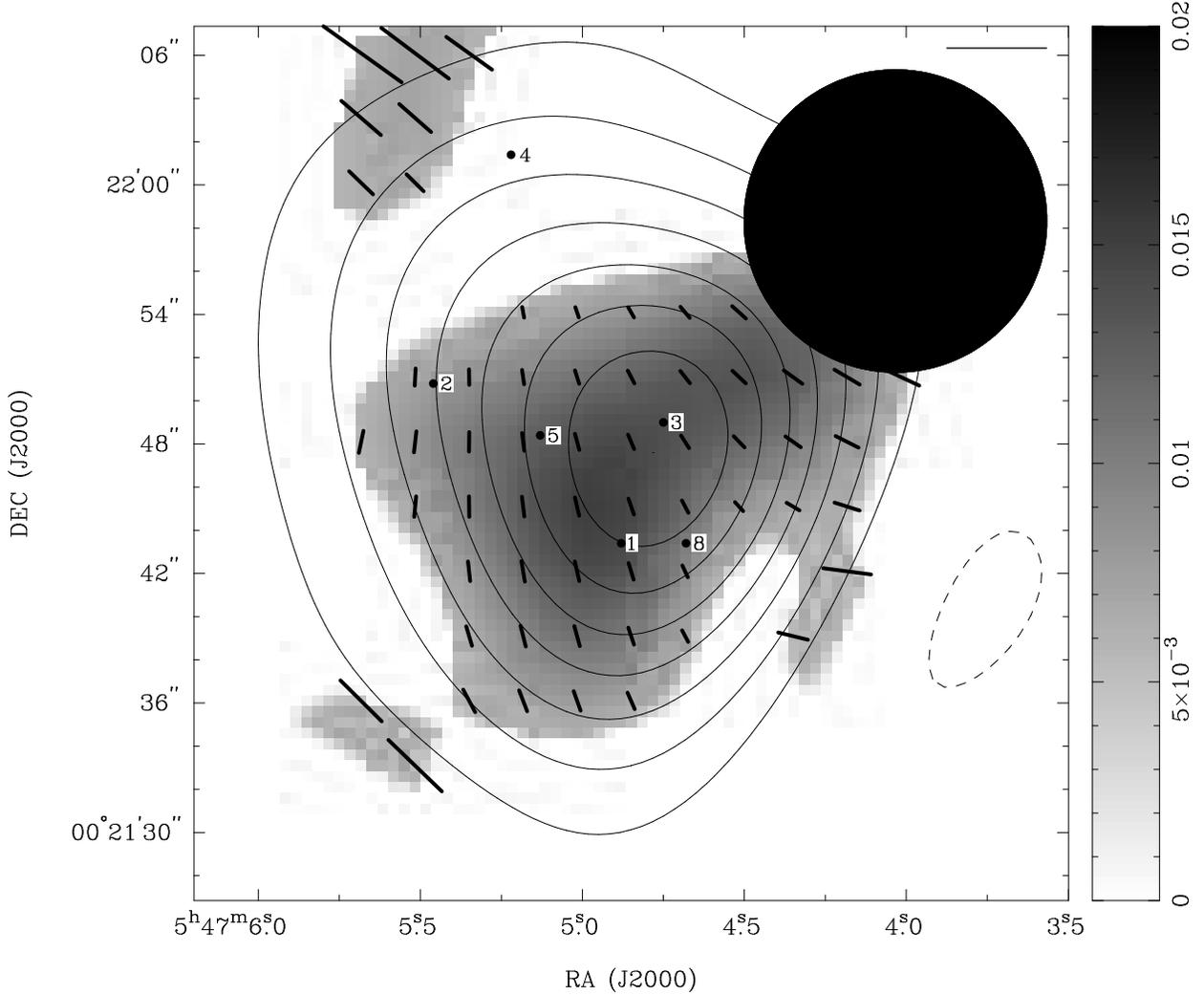}
\epsscale{0.4}
\caption{
Convolution of Figure~\ref{NGC2071con} with
a gaussian beam of 14$^{\prime \prime}$ FWHM.
The contours represent
the Stokes I emission at -0.13, 0.13, 0.26, 0.39, 0.52, 0.65, 0.78, and 0.92
Jy beam$^{-1}$.
The pixel gray scale show $3\sigma$ polarized
intensity ($\sqrt{Q^{2} + U^{2}}$) measured in Jy beam$^{-1}$, for
the dust continuum emission. The black line segments
are the polarized dust emission map. The fractional polarization
is represented by the length of the line and the P.A. by the 
orientation of the line segment.
The scale for the line segments is given by the bar at the
upper right corner which represents 0.1 in fractional polarization.
}
\label{NGC2071convolved}
\end{figure}

Under most conditions it is assumed that dust grains will be aligned
in the presence of a magnetic field. Using this premise \citet{Matthews2002}
inferred a magnetic field morphology for NGC2071IR which, they
concluded, does not show the expected hourglass shape predicted
by theory for a magnetically supported cloud.
This conclusion seems to be in agreement with  the
Stokes I emission present in their maps, which does not
show flattening.
In order to fully
compare with \citet{Matthews2002} we convolved our
map with a Gaussian beam of $\sim$ 14$^{\prime \prime}$ at FWHM;
the convolved map is shown in Figure~\ref{NGC2071convolved}. The
lower resolution Stokes I image has all traces of flattening erased,
agreeing in morphology with the contour map shown by
 \citet{Matthews2002}. In the same way, the polarization map
shows a fairly uniform profile over the core with a mean P.A. of
43.6$^\circ \pm$ 4.1$^\circ$, while the JCMT data has, 
over the same box, a mean P.A. of 36.9$^\circ \pm$ 3.5$^\circ$ 
(taken from Table 1 in \citet{Matthews2002}).
We see that the sudden change in polarization P.A.
in the central region of the core (shown in  Figure~\ref{NGC2071con})
is completely lost in the lower resolution map.

Unlike the lower resolution data, our higher resolution BIMA map
shows an elongated core and (if the polarization is interpreted as
being due to magnetic field alignment of grains),
a magnetic field that is parallel to
the major axis of the core in the north, approximately orthogonal
at the center, and about $45^{\circ}$ to the major axis
in the south of the core. This projected field morphology would
require
an abrupt $90^{\circ}$ twist of the field from the north to the center.
The field would then be along the minor axis at the center of the core,
in agreement with a strong magnetic field model with contraction
along field lines. However, at both the northern and southern edges of
the core, and over the more extended region mapped with SCUBA, the
field would be twisted up to $\sim 90^{\circ}$.

An alternative explanation
for this polarization pattern is that grains are mechanically aligned by
the powerful molecular outflow directed along the magnetic field at
a P.A. $\sim 40^{\circ}$.
\citet{Rao1998} observed
dust polarized emission from Orion KL using the BIMA telescope
at 1.3 mm. They noticed
an abrupt change of 90$^{\circ}$ in P.A. from one point to another
in their polarization data (similar to our case),
which they found difficult to explain
in terms of a twisting magnetic field. They concluded that this
was evidence for mechanical alignment by the powerful
 outflow of Orion KL.  In our case, the relative alignment of the polarization
pattern in Figure~\ref{NGC2071con} with the outflow direction suggests
mechanical alignment over most of the core (and over the more 
extended region mapped by SCUBA), while at the center magnetic
alignment would yield a magnetic field direction consistent with the
orientation of the outflows powered by IRS 1 and IRS 3. The 
different efficiencies of the alignment mechanism may be explained
by an outflow which is not dynamically significant at the
center of the core (between ISR 1 and IRS 3). This scheme seems
to be consistant with the radiative torques model at similar densities
\citep{Cho2005}.

We follow the formulation described by \citet{Mooney1995,Mezger1990} to
calculate the column density and mass. Both quantities are give by the
following equations,

\begin{equation}
\label{cd}
N_{\rm{H}}/{\textnormal{cm}}^{-2}=1.93 \times 10^{15}
\frac{(S_{\nu,int}/\rm{Jy})
\lambda^{4}_{\mu \rm{m}}}
{(\theta_{s}/\rm{arcsec})^{2}(Z/Z_{\sun})bT}\frac{e^{x} - 1}{x}
\end{equation}

\begin{equation}
\label{mass}
M_{\rm{H}}/\textnormal{M}_{\sun}=4.1 \times 10^{-10}
\frac{(S_{\nu,int}/\rm{Jy})\lambda^{4}_{\mu \rm{m}}
 D^{2}_{\rm{kpc}}} {(Z/Z_{\sun})bT} \frac{e^{x} - 1}{x},
\end{equation}

\noindent where $S_{\nu ,int}$ is the total flux from the source,
$\theta_{s}=\sqrt{\theta_{s,min} \times \theta_{s,max}}$ is the 
angular source size, $Z/Z_{\sun}=1$ is relative metallicity, T is the
dust temperature, $b=3.4$ reproduces cross sections estimates
for dust around deeply embedded {\bf IR} sources \citep{Mezger1990}, 
$x=\frac{1.44 \times 10^{4}}
{\lambda_{\mu \rm{m}}T}$ 
is the $\frac{hc}{\lambda kT}$ factor for the Planck
function, and $D_{\rm{kpc}}$ is the distance to the source in kpc.
Using the obtained total flux ($S_{\nu,int}=1.87$ Jy) we calculated
$N_{\rm{H}}=1.5 \times 10^{23} \textnormal{cm}^{-2}$ and
$M_{\rm{H}}=2.8$  $\textnormal{M}_{\sun}$. 
Using a $30^{\prime \prime}$ angular distance over the source, 
which is equivalent to $1.75 \times 10^{17} \textnormal{cm}$ 
at a distance of 390 pc, we obtained a volume density of
$n_{\rm{H}}=1.25 \times 10^{6} \textnormal{cm}^{-3}$. Interstellar
dust has been argued to radiate polarized emission only at low 
$A_{\rm{v}}$ \citep{Goodman1995,Lazarian1997c}.
However, using the relation between
$A_{\rm{v}}$ and 
$N_{\rm{H}}$ given by \citet{Bohlin1978}

\begin{equation}
\left< N(\rm{H}) / A_{\rm{v}} \right> = 5.8 \times
10^{21} \textnormal{cm}^{-2} \textnormal{mag}^{-1},
\end{equation}

\noindent we find $A_{\rm{v}}=26$ $\textnormal{mag}$, 
which is greater
than the upper limit, $1 < A_{\rm{v}} < 10$ mag,
 proposed by \citet{Lazarian1997c}, which suggests that dust
grains will align at higher densities than previously thought.

\citet{Houde2001} observed NGC2071IR and concluded that
the magnetic field  is aligned with the outflow
by the comparison of widths of spectral lines
of neutral and ionized species. This picture agrees with the one we
discuss here. However, they did not provide
direct (polarization) evidence of a magnetic field in the NGC2071IR outflow.
Figure~\ref{NGC2071co} shows CO $J=2 \rightarrow 1$ polarized emission maps
from the outflow. Line polarization is predicted in the presence
of a magnetic field \citep{Goldreich1981}; therefore, this is direct
evidence for a magnetic field in the outflow of NGC2071IR.
The CO polarization has orthogonal position angles at different locations
in both the red and blue line wings. Unfortunately, interpretation is
ambiguous, for
there is a $90^{\circ}$ ambiguity between the predicted polarization P.A. and
the magnetic field, and the direction of line polarization can flip
by $90^{\circ}$ within a region \citep{Goldreich1981}. The resolution of the
ambiguity, in the general case, requires a more detailed study of the region 
(e.g., \citet{Cortes2005}). However, in the case of NGC2071IR outflow
the coincidence between one of the polarization orientations and 
the outflow P.A. suggests that right interpretation is indeed, a
magnetic field along the outflow.
Additionally, it is also
known that CO $J=2 \rightarrow 1$ polarized emission will generally trace
the magnetic field at a lower density from polarized dust emission,
which raises the question of how the fields in the core and outflow
are connected. 

In summary, we suggest that the
magnetic field morphology in NGC2071IR is best described as a magnetic
field aligned with the bipolar outflow which would be perpendicular
to the major axis of the core in Figure~\ref{NGC2071con}.
This is consistent with a flattening of the core along magnetic field
lines and allows a connection between the field in the outflow and
in the core.  Mechanical alignment for grains at
the northern-west and southern-east edges of the map (relative to the peak in Figure~\ref{NGC2071con}),
which will produce polarization parallel
to the magnetic field \citep{Lazarian1994,Lazarian1997a},
would explain the observed polarization morphology. However,
this may  require an
outflow generated by both sources (IRS 1 and IRS 3), which
may be possible. Also, it is important to consider that the outflow 
and the polarized dust emission might have
different orientations with respect to the line of sight, 
even if their projections on the plane of the sky
agree in orientation. 
Near the center of the core the direction of dust polarized
emission cannot be explained by mechanical alignment; a possible
explanation would be magnetic alignment winning over a dynamically weak
outflow in that part of the core, which is consistent with a magnetic
field parallel to the outflow at center of the core.

\section{Summary and conclusions}

We observed NGC2071IR and successfully detected
 CO $J=2 \rightarrow 1$ line and 1.3 mm dust continuum polarized emission
with a resolution of 4$^{\prime \prime}$.

We found direct evidence for a magnetic field in the outflow
of NGC2071IR through CO $J=2 \rightarrow 1$ polarized line emission.
Also, from polarized dust emission, we suggest a magnetic field
in the core along its minor axis and parallel to the outflow
direction, which is consistent with the observed flattening of the
core. Over most of the region the polarization is parallel to
the outflow suggesting mechanical grain alignment. This interpretation
provides a consistent picture for the field morphology.

We also estimated a visual extinction $A_{\rm{v}}=26$ mag 
relative to
a column density of $N_{\rm{H}}=1.5 \times 10^{23}$ cm$^{-2}$.
This result suggests that the dust will polarized
efficiently at greater densities than previously thought.

This research was partially funded by NSF grants AST 02-05810 and 02-28953.
The BIMA array was operated with support from the National Science
Foundation under grants AST-02-28963 to UC Berkeley, AST-02-28953 to 
U.\ Illinois, and AST-02-28974 to U.\ Maryland.
BCM acknowledges funding a postdoctoral fellowship from the Natural
Sciences and Engineering Research Council of Canada. 

\bibliography{biblio}

\newpage

{\normalsize \begin{deluxetable}{ccc}
\tablecolumns{4}
\tablewidth{0pc}
\tablenum{1}
\tablehead{
\colhead{Offsets in arcsec} & 
\colhead{$P_{Dust}$} &
\colhead{$\phi_{Dust}$} 
}
\tablecaption{Fractional polarization and 
position angle for dust
polarization observations from NGC2071IR. 
Data were interpolated at a tolerance of 1$^{\prime \prime}$
 that corresponds to approximate 1.9 $\times$
10$^{-3}$ pc using a distance to NGC2071 of 390 pc.}
\startdata
(0,-6.0)    & 0.05$\pm$0.01 & 22.9$\pm$7.4 \\
(2.0,-4.0)  & 0.1$\pm$0.03  & 16.6$\pm$8.4   \\
(0,-4.0)    & 0.05$\pm$0.01 & 16.6$\pm$7.6 \\
(-4.0,-6.0) & 0.23$\pm$0.09 & -80.3$\pm$9.2 \\
(-2.0,-2.0) &0.03 $\pm$0.01 &-17$\pm$8.9 \\
(-3.0,-2.0) & 0.03$\pm$0.01 & -16.5$\pm$8.8 \\
(-6.0,2.0)  &0.06$\pm$0.02  & 65$\pm$9 \\
(-4.0,4.0)  &0.07$\pm$0.02  & 60.7$\pm$7.4 \\
(-6.0,4.0)  &0.07$\pm$0.02  & 66.1$\pm$7.3 \\
(-4.0,6.0)  &0.09$\pm$0.02  & 57.6$\pm$7.6 \\
(5.0,18.0)  &0.3 $\pm$0.1   & -4.6$\pm$9.3 \\
(3.0,18.0)  &0.2$\pm$0.08   & -4.2$\pm$9.2 \\
\enddata
\label{Table1.3}
\end{deluxetable}
}

\newpage

\end{document}